\documentclass[10pt,prb,aps,twocolumn,superscriptaddress,showpacs,floatfix]{revtex4}

\usepackage{graphicx}
\usepackage{amssymb}

\def\be{\begin{equation}}
\def\ee{\end{equation}}
\def\ba{\begin{eqnarray}}
\def\ea{\end{eqnarray}}

\begin{document}

\title{Quantum phase transitions in disordered dimerized \\
quantum spin models and the Harris criterion}

\author{Dao-Xin Yao}
\affiliation{State Key Laboratory of Optoelectronic Materials and Technologies, School of Physics and Engineering,
Sun Yat-sen University, Guangzhou 510275, China}

\author{Jonas Gustafsson}
\affiliation{Department of Physics, Boston University, 590 Commonwealth Avenue, Boston, Massachusetts  02215, USA}
\affiliation{Theoretical Physics, Royal Institute of Technology, SE-10691 Stockholm, Sweden}

\author{E.~W.~Carlson}
\affiliation{Department of Physics, Purdue University, West Lafayette, Indiana 47907, USA}

\author{Ander W.~Sandvik}
\affiliation{Department of Physics, Boston University, 590 Commonwealth Avenue, Boston, Massachusetts  02215, USA}

\date{\today}

\begin{abstract}
We use quantum Monte Carlo simulations to study effects of disorder on the quantum phase transition occurring versus the ratio 
$g=J/J'$ in square-lattice dimerized $S=1/2$ Heisenberg antiferromagnets with intra- and inter-dimer couplings $J$ and $J'$.
The dimers are either randomly distributed (as in the classical dimer model), or come in parallel pairs with horizontal or 
vertical orientation. In both cases the transition violates the Harris criterion, according to which the correlation-length 
exponent should satisfy $\nu\ge 1$. We do not detect any deviations from the three-dimensional O$(3)$ universality class 
obtaining in the absence of disorder (where $\nu \approx 0.71$). We discuss special circumstances which allow $\nu<1$
for the type of disorder considered here.
\end{abstract}


\maketitle

Studies of effects of disorder (randomness) at phase transitions have a long history in statistical physics, with the celebrated 
``Harris criterion'' providing a guide for when disorder should be expected to be relevant, i.e., leading to changes in the critical 
exponents.\cite{harris74} The statement by Harris is that for a $d$-dimensional classical system, the exponent $\nu$ governing the 
divergence of the correlation length should satisfy $\nu \ge 2/d$ in the presence of disorder. If the exponent for an unperturbed 
(clean) system does not satisfy this relationship, then, in the presence of disorder, if the transition remains well defined 
(i.e., it is not smeared, with different transition points in different regions of the system \cite{vojta04}) a new universality 
class should obtain in which the relationship does hold. Alternatively, in some quantum systems the behavior instead 
becomes activated, with exponential scaling instead of power-law singular behavior.\cite{fisher92}

The Harris criterion was originally derived based on a natural assumption of how the local critical temperature in some region 
of a classical system with random couplings can be directly related to local fluctuations in the average coupling strength (or impurity 
concentration). Consistency with a single critical temperature for the whole system (no smearing) then leads to the requirement 
$\nu \ge 2/d$.\cite{harris74} This condition was later re-derived using an alternative, more rigorous method, which allowed for extension 
to some quantum systems as well.\cite{chayes86} The effective statistical-mechanics problem for a quantum system at temperature $T=0$ 
corresponds, through the Euclidean path integral, to a classical system in $d+1$ dimensions (under the assumption that quantum mechanical 
effects due to Barry phases can be neglected, which is not always the case \cite{sachdevbook}). Since disorder is introduced only in the 
original spatial dimension (corresponding to columnar disorder in the $d+1$ dimensional classical system), the dimensionality to use in the 
Harris criterion for a quantum system is presumed to be just $d$, not $d+1$.\cite{chayes86} Hence, for two-dimensional quantum 
spin systems, which we will study in this paper, one would expect $\nu \ge 1$ at a quantum phase transition in the presence of disorder.

 %------------------------------------------------------------------------------------------
\begin{figure}[b]
\centerline{\includegraphics[width=7cm, clip]{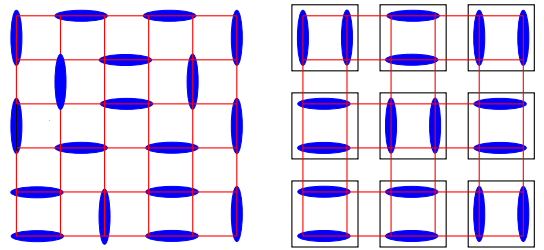}}
\vskip-2mm
\caption{(Color online) Dimerized systems with two types of configurational disorder. The dimers (shown as ovals) 
are spin pairs with 
interactions $J$ stronger than the inter-dimer couplings $J'$. In the random dimer model (left) all close-packed dimer 
configurations are included, whereas in the random plaquette system (right) a superlattice of $2\times 2$ plaquettes has horizontal
or vertical dimer pairs within the plaquettes.}
\label{fig1}
\vskip-3mm
\end{figure}
%------------------------------------------------------------------------------------------

We will discuss quantum phase transitions in spin-$1/2$ dimerized Heisenberg antiferromagnets on the square lattice. A dimer consists 
of two nearest-neighbor spins coupled by a Heisenberg interaction of strength $J$. All spins belong to exactly one dimer, of which 
there are $N/2$ for a lattice with $N=L\times L$ sites and $L$ even (and we use periodic boundary conditions). The dimers are coupled 
to each other through all the other nearest-neighbor bonds, with weaker coupling $J'$. The Hamiltonian is thus
\begin{equation}
H= J \sum_{\langle i,j\rangle }\mathbf{S}_i \cdot \mathbf{S}_j +J' \sum_{\langle i,j\rangle '}
\mathbf{S}_i \cdot \mathbf{S}_j,      
\label{hamilton}
\end{equation}
where $\langle i,j\rangle$ is the set of dimers and $\langle i,j\rangle '$ denotes the rest of the nearest-neighbor pairs. We will 
here consider disorder in the form of random configurations of the dimers, constructed in two different ways as illustrated in 
Fig.~\ref{fig1}.

For a system with a regular (non-random) dimer patterns, there is quantum phase transition as a function of the coupling ratio $g=J/J'$. 
Well studied examples include dimers arranged in columns \cite{matsumoto01,wenzel08} or between the layers of a 
bilayer.\cite{sandvik94} According to standard symmetry considerations the phase transition should be in the universality class of the 
O$(3)$ (classical Heisenberg) model with $d=3$. There are, however, subtleties related to Berry phases and the way the continuum limit is 
taken in effective field theories, such as the (2+1)-dimensional non-linear $\sigma$-model.\cite{chn,read89,einarsson91,chubukov94} Large 
scale quantum Monte Carlo (QMC) calculations of bilayers and columnar dimers have given critical exponents in very good agreement with 
the expected universality class. Results for some other patterns, e.g., staggered dimers (where every second row of a columnar dimer 
pattern is shifted by one lattice spacing) are currently puzzling, with either a different universality class obtaining \cite{wenzel08} 
or unexpectedly large corrections to scaling.\cite{jiang09,wessel10}

Disorder can be introduced in these dimerized systems in many different ways. We are here interested in systems with maintained SU$(2)$
symmetry. Since $\nu < 1$ ($\approx 0.71$),\cite{campostrini02} disorder is expected to be relevant by the Harris criterion.  On possibility 
is to dilute the system by removing a fraction of the spins at random. In general this will completely destroy the phase transition, however, 
because in the non-magnetic phase the removal of a single spin leads to an uncompensated magnetic moment (the remaining spin of the dimer with a 
vacancy). For a finite concentration of vacancies the subsystem of liberated moments exhibits long-range order. One can circumvent this problem 
by removing whole dimers.\cite{sandvik02a,vajk02} A large-scale Monte Carlo study of an effective 3-dimensional classical model corresponding 
to this situation indicated a generic transition (at fixed dilution below the classical percolation threshold) satisfying the Harris 
criterion.\cite{vojta06}

Here we investigate the transition in the presence of two different types of {\it configurational} disorder, illustrated in Fig.~\ref{fig1}.
In the {\it random dimer model} (RDM), we average over the ensemble of all possible dimer configurations, as in the classical dimer model. 
In the {\it random plaquette model} (RPM), we subdivide the lattice into $2\times 2$ plaquettes and place two parallel dimers within all the 
plaquettes. Starting from a clean system of all horizontal dimers (a columnar configuration), we rotate a fraction $p$ of the dimer pairs 
by $90^\circ$. In this case $p$ is a well defined measure of the degree of disorder in the system, with maximum disorder 
at $p=1/2$ (which is the case we consider here, unless otherwise stated). In the RDM, on the other hand, there is no 
tunable impurity concentration or disorder strength. In addition, in this case the disorder is correlated, as the averaged dimer-dimer 
correlations decay as $1/r^2$ in the close-packed dimer system.\cite{fisher61} The prerequisites of the Harris criterion may then be violated 
in the RDM.\cite{cruz86} I contrast, the RPM dimers are only locally correlated (within the individual plaquettes), which should be of no 
relevance in a coarse-graining sense. Our objective here is to investigate the role of correlated disorder in the RDM and to test the validity 
of the Harris criterion in both models.

We have performed quantum Monte Carlo simulations using the stochastic series expansion (SSE) method.\cite{sandvik99} Sufficiently low 
temperatures are used for obtaining ground state results for lattices with $L$ up to $40$ (using  procedures for checking the $T\to 0$ 
convergence discussed in Ref.~\onlinecite{sandvik02b}). We will discuss finite-size scaling of several quantities. The staggered structure 
factor is defined as 
\begin{equation}
S(\pi,\pi)=\frac{1}{N}\left \langle \left ( \sum_{i=1}^N S^z_i\phi_{i}\right )^2 \right \rangle  = N\langle m_s^2\rangle
\end{equation} 
where $\phi_i=\pm 1$ is the staggered phase factor and $m_s$ is the sublattice magnetization. $S(\pi,\pi)$ should
scale at a $d=2$ quantum-critical point as $L^{z-\eta}$, where the exponent $\eta \approx 0.037$ in the O(3) universality class \cite{campostrini02} and 
the dynamic exponent $z=1$. If the universality class changes due to the disorder, the new exponents are not known. The Binder ratio,
\begin{equation}
Q_2 = \frac{\langle m_s^4\rangle}{\langle m_s^2\rangle^2},
\end{equation}
is a dimensionless quantity with no size corrections (asymptotically) at criticality; 
$Q_2(L) \to {\rm constant}$ at the critical point. We also study 
the spin stiffness, the second derivative of the ground state energy $E(\phi)$ (per spin) in the presence of a boundary phase twist $\phi$;
\begin{equation}
\rho_s=\frac{\partial^2E(\phi)}{\partial \phi^2},
\end{equation}
which is obtained in the SSE simulations in the standard way using winding number fluctuations.\cite{sandvik97} It's scaling at
criticality is only governed by the dynamic exponent $z$; $\rho_s \sim L^{-z}$ in two dimensions.

We study disorder-averaged quantities. For each system size, at least several hundred configurations were used. We apply the standard
finite-size scaling formalism, according to which a quantity $A$ should depend on the lattice length $L$ and the deviation from the critical 
point $g_c$ according to
\begin{equation}
A(g,L)=L^{\kappa}(1+aL^{-\omega})f[(g-g_c)L^{1/\nu}],
\end{equation} 
where $f$ is a non-singular function, $\nu$ the correlation-length exponent, and $\kappa$ depends on the quantity considered, as 
discussed above. Here we have also included a subleading correction $(1+aL^{-\omega})$, which in some cases is needed in order to obtain good 
fits to the data. 

\begin{figure}[t]
\centerline{\includegraphics[width=7.5cm, clip]{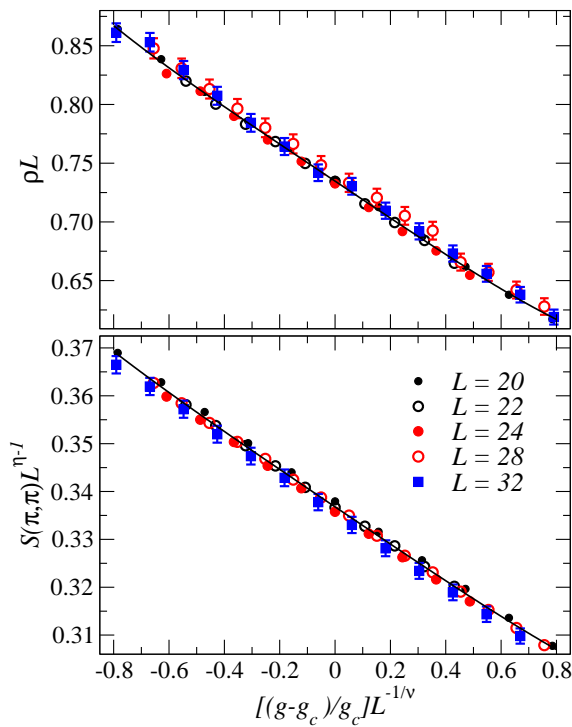}}
\vskip-2mm
\caption{(Color online) Finite-size scaling of the spin stiffness (top panel) and the staggered structure factor (bottom panel) 
of the RDM, using the $d=3$ Heisenberg exponents ($\eta=0.0375$, $\nu=0.7115$) and $g_c=2.145$. Where not shown, the error bars are smaller
than the symbols. Note that the statistical errors for a given lattice size $L$ are correlated, because the same random dimer 
configurations were used for all coupling rations $g$ and the sample-to-sample fluctuations are larger than the QMC statistical
errors.}
\label{fig2}
\vskip-3mm
\end{figure}

\begin{figure}[t]
\centerline{\includegraphics[width=7.5cm, clip]{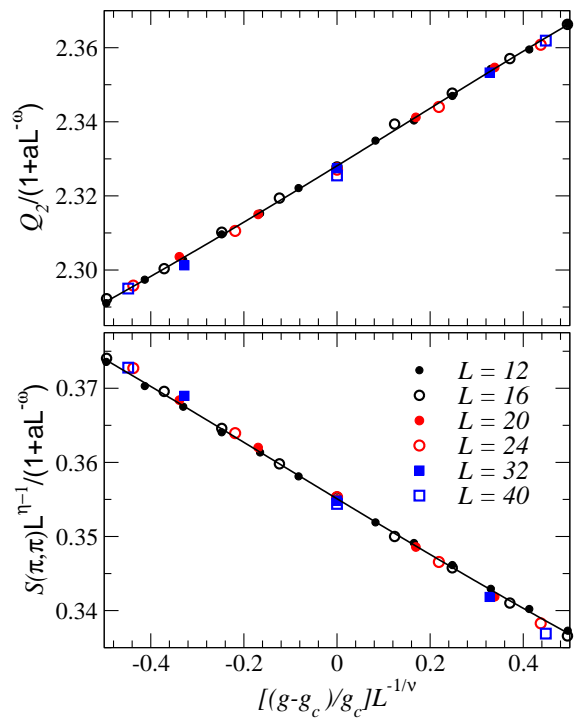}}
\vskip-2mm
\caption{(Color online) Finite-size scaling of the Binder ratio (top panel) and the staggered structure factor (bottom panel) or the RPM, 
using the $d=3$ Heisenberg exponents ($\eta=0.0375$, $\nu=0.7115$) and the critical point $g_c=1.990$. The subleading exponent $\omega \approx 1$ 
in both cases and the prefactor $a\approx -0.5$ for $Q_2$ and $a\approx -0.1$ for $S(\pi,\pi)$. The error bars are at most of the order of 
the size of the symbols.}
\label{fig3}
\vskip-3mm
\end{figure}

We have analyzed SSE
data for the RPM and RDM in different ways, with and without scaling corrections, keeping all the exponents and the critical coupling 
ratio $g_c$ as variables in the fitting procedure or keeping some of them fixed to values obtained in other fits, using different sets
of system sizes, etc. In all cases we find that $z=1$ can describe the data very well, and therefore conclude that the dynamic exponent is 
not changed by the disorder. Surprisingly, all other exponents, as well, come out very close to their $d=3$ Heisenberg values. In a final stage 
of the analysis we therefore fix all the exponents to their best available $d=3$ $O$(3) values \cite{campostrini02} and only adjust $g_c$ to 
optimize the fits. We consistently obtain good fits with values of $g_c$ that agree among the different quantities studied; our final estimates 
for the critical coupling ratios are $g_c=2.145 \pm 0.001$ (RDM) and $g_c=1.990 \pm 0.001$ (RPM). Figs.~\ref{fig2} and \ref{fig3} show some 
examples of data fits; for $\rho_s$ and $S(\pi,\pi)$ of the RDM without subleading corrections in Fig.~\ref{fig2}, and for $Q_2$ and 
$S(\pi,\pi)$ of the RPM with subleading corrections in Fig.~\ref{fig3}.

 %------------------------------------------------------------------------------------------
%------------------------------------------------------------------------------------------

 %------------------------------------------------------------------------------------------
%------------------------------------------------------------------------------------------

The conclusion of this study is, thus, that the transitions in both the RPM and RDM violate the Harris criterion. It has been pointed out 
before that this criterion, in fact, contains several implicit assumptions that may make it inapplicable (or require extensions) for 
some systems.\cite{cruz86,pazmandi97} In addition, the criterion should really be written as $\nu_{\rm FS} \ge 2/d$,\cite{chayes86,pazmandi97} 
where the finite-size correlation-length exponent $\nu_{\rm FS}$ is exactly the one extracted in scaling procedures such as those we have used 
above. The intrinsic correlation length can be detected using a modified procedure \cite{pazmandi97} involving scaling relative to individual 
finite-size sample definitions of the critical point. The fact that our result shows unambiguously that $\nu_{\rm FS}<2/d$ implies 
\cite{pazmandi97} that the sample-to-sample fluctuations of the critical point are smaller than assumed in the original derivations of the 
Harris criterion. We have studied these fluctuations and, indeed, find that they are very small (in fact, so small that it is difficult to 
study their size dependence quantitatively). The modified scaling procedure therefore also produces results consistent with the same 
$O$(3) exponents. It seems, therefore, that these exponents also are the intrinsic exponents, $\nu_{\rm FS}=\nu$ and all exponents have their 
clean-system values.

A transition in violation of $\nu_{\rm FS}\ge 2/d$ and unchanged exponents have also been found in the $d=2$ disordered bosonic Hubbard model, at 
the special multi-critical point at the tip of the Mott lobes.\cite{kisker97} There it was argued \cite{pazmandi98} that the the critical point 
does not depend on the disorder strength, which violates the prerequisite of the Harris criterion of the possibility to drive the transition by 
tuning the disorder strength \cite{pazmandi98} (although this is called into question by recent work \cite{pollet09}). This is the case 
also for our RDM, where there is no notion of disorder strength or concentration. In the RPM, there is, however, a clearly observable 
dependence on the probability $p$ characterizing the ratio of horizontal and vertical dimer pairs. At $p=0,1$, the critical value is 
the smallest, $g_c=1.909$,\cite{matsumoto01,wenzel08} and the maximum value is $g_c=1.990$ at 
$p=1/2$, as reported above. The curve $g_c(p)$ is symmetric about the point $p=1/2$ that we analyzed above, and the local dependence on $p$ is 
particularly small there (but we do not know the exact form of $g_c$ versus $p$), which may explain the smallness of the sample-to-sample 
fluctuations in $g_c$ (which, according to Ref.~\onlinecite{pazmandi97} can account for $\nu_{\rm FS}< 2/d$). On the other hand, 
we have also studied $p=1/4$ and also there find no changes in the exponents.

Here the recent ``inclusion theorem'' by Pollet {\it et al.} \cite{pollet09} should be noted. At first sight it (and similar standard arguments
for Griffiths-McCoy phases \cite{sachdevbook}) appears to rule out a direct transition between the N\'eel state and the featureless gapped phase 
in the models considered here. One could argue that, in the non-magnetic phase close to $g_c$, one could always find (in the thermodynamic limit) 
infinitely large regions in the N\'eel phase. The system as a whole would then not be gapped. The RDM is not, however, amenable to the 
analysis of Ref.~\onlinecite{pollet09}, because the disorder is constrained and correlated, not given by just a local distribution. The disorder 
in the RPM is given by a local distribution, but $p=1/2$ gives the extremal $g_c$ and, thus, is also not covered by the theorem (the proof of which 
specifically excludes such extremal points). For generic $p$ the theorem should apply, and there may then be other aspects of the transition not 
considered here, e.g., an intervening gapless phase with no long-range magnetic order or a smeared transition. We have not observed any such behaviors 
at $p=1/4$, but it is possible that much larger systems have to be studied in order to reveal effects of rare regions. 

As already noted, the RDM is a special case in another way, too. The constrained disorder of close-packed dimers leads to dimer-dimer correlations 
decaying as $1/r^2$.\cite{fisher61} This represents the border-line case of disorder correlated according to a power-law $1/r^a$, where for 
$a>2$ the usual Harris criterion should apply (in cases where the criterion is valid for uncorrelated disorder), and for $a<2$ a modified 
criterion was presented. \cite{weinrib83} This may be of no relevance here, however, since the usual Harris criterion is not valid for the 
uncorrelated RPM. 

In summary, we have studied configurational disorder in dimerized square-lattice $S=1/2$ Heisenberg models. We find no change of universality 
class of the N\'eel to nonmagnetic quantum phase transition, in violation of the Harris criterion. While this criterion does not state
the fixed point to which the disordered system flows, this point should, if the criterion is valid, satisfy $\nu_{FS}>1$ for dimensionality
$d=2$, which is ruled out by our results. Our study reinforces the notion that the Harris criterion can be violated.\cite{pazmandi97} The 
transition does not represent the most likely scenario discussed  in Ref.~\onlinecite{pazmandi97}, where the exponents still would change 
due to the disorder. While such a case of unchanged exponents with $\nu_{FS}<1$ has been claimed in at least one past study,\cite{kisker97} our 
model is particularly appealing from a numerical perspective, because subtleties related to other intervening phases \cite{pollet09} 
do not apply for the particular types of disorder we have used here. It would still be interesting to study the RPM in more detail for 
different values of the dimer-orientation parameter $p$, where issues relating to rare fluctuations should come into play for $p\not=1/2$.\cite{pollet09} 
Our results so far do not show any evidence for a glassy intermediate phase between the N\'eel and nonmagnetic states.

It is also interesting to note that we do not observe any anomalous scaling behavior, which has recently 
been discussed in certain regularly dimerized systems.\cite{wenzel08,jiang09} This could in part be explained by the fact that the 
numerical precision we have obtained here for disordered systems is not as high as in the studies of clean systems. One would at least 
naively expect potential effects of uncompensated Berry phases \cite{sachdevbook} to be larger in systems with random dimer arrangements.

DXY acknowledges support from Sun Yat-sen University, EWC by Research Corporation and NSF Grant No. DMR 08-04748, and AWS
by the NSF under Grant No.~DMR-0803510.

\null

\end{document}